\documentclass{article}

\usepackage{arxiv}
\usepackage[utf8]{inputenc} 
\usepackage[T1]{fontenc}    
\usepackage{hyperref}       
\usepackage{url}            
\usepackage{booktabs}       
\usepackage{amsfonts}       
\usepackage{nicefrac}       
\usepackage{microtype}      
\usepackage{lipsum}		
\usepackage{graphicx}
\usepackage{natbib}
\usepackage{doi}
\usepackage{amsmath,amsthm,amssymb}
\usepackage{babel}
\usepackage{blindtext}
\usepackage{subfiles} 
\usepackage{dsfont}
\usepackage{bm}
\usepackage[ruled,linesnumbered,noend]{algorithm2e}
\usepackage{mathtools}
\usepackage{indentfirst}
\usepackage{fancyhdr}
\usepackage{mathabx}
\usepackage{booktabs} 
\usepackage{caption} 
\usepackage{setspace} 
\usepackage{lscape}
\usepackage{pdfpages}
\usepackage{float}
\usepackage[nottoc,numbib]{tocbibind}
\usepackage{amsfonts}
\usepackage{amsmath}
\usepackage{eucal}

 \usepackage[pagewise]{lineno}

\usepackage{hyperref}
\title{Scalable Bayesian bi-level variable selection in generalized linear models}
\author{{Younès Youssfi}\\
	ENSAE Paris\\
	Institut Polytechnique de Paris\\
	91120 Palaiseau, France\\
	\texttt{younes.youssfi@ensae.fr}\\
	\And
	{Nicolas Chopin\thanks{Corresponding author}} \\
	ENSAE Paris\\
	Institut Polytechnique de Paris\\
	91120 Palaiseau, France\\
	\texttt{nicolas.chopin@ensae.fr}\\
}

\date{}
\renewcommand{\shorttitle}{Scalable Bayesian bi-level variable selection in generalized linear models}

\begin{document}
\maketitle

\begin{abstract}
    Motivated by a real-world application in cardiology, we develop an
    algorithm to perform Bayesian bi-level variable selection in a generalized
    linear model, for datasets that may be large both in terms of the number of
    individuals and the number of predictors. Our algorithm relies on the
    waste-free SMC (Sequential Monte Carlo) methodology of \cite{wastefreeSMC},
    a new proposal mechanism to deal with the constraints specific to
    bi-level selection (which forbid to select an individual predictor if its
    group is not selected), and the ALA (approximate Laplace approximation)
    approach of \cite{rossell2021approximate}. We show in our numerical study
    that the algorithm may offer reliable performance on large datasets within
    a few minutes, on both simulated data and real data related to the
    aforementioned cardiology application.
    
\end{abstract}


\section{Introduction}\label{sec:intro}

\subsection{Motivation}\label{sub:motivation}


While useful more generally, the approach developed in this paper was
initially motivated by a public health dataset recording the medical
history of a large number of individuals that may or may not have
suffered from sudden cardiac death (SCD); this dataset will be
described more fully later. One may use this data to determine whether
consumption of medical drugs or hospitalization may increase the odds of an SCD event.
Unfortunately, the number of potential drugs and diseases is very large, and their
incidence in the studied population vary a lot. This makes it
difficult to assess the impact of drugs and diseases that are rarely prescribed or observed. On
the other hand, there are official nomenclatures for drugs and diseases, which
can be classified into groups with similar properties. Hospital diagnoses are coded according to the International
Classification of Diseases and drugs are coded according to the Anatomical Therapeutic Chemical system, that classifies them according to the organ or system on which they act and their therapeutic, pharmacological,
and chemical properties. Therefore, there is clear medical interest in
determining automatically whether there is enough information in the
data to indicate that a particular drug or disease affects SCD, or, if not, whether the
group it belongs to does. 

This led us to develop a bi-level variable selection procedure,
based on a binary regression (outcome variable is whether the individual
had an SCD event) model, and which should work reliably for a fairly
large number of individuals, variables and groups. In
addition, we wanted this procedure to be Bayesian, in order to be able
to obtain posterior probabilities of inclusion (rather than simply 0/1
answers).

There are surprising few papers on Bayesian bi-level variable selection, and
most of them focus on linear regression with Gaussian noise \citep{MR3533632,
MR3612612, MR4085862}. For such a model, one may integrate out the regression
coefficients (the prior provided is Gaussian) to obtain the marginal posterior
distribution over a finite space (the inclusion of either individual variables
or groups). Even so, designing a MCMC able to efficiently explore that finite
space is challenging. Such discrete distributions tend to exhibit strongly
separated modal regions, and a MCMC chain may fail to escape one of this
region. We refer in particular to the numerical experiments of
\cite{schafer2011sequential} that show that various MCMC schemes  may lead to
unstable estimates because of this problem. Of course, this issue gets worse
when the number of variables increases, making MCMC unable to scale properly
with datasets with a large number of variables (and groups).

\subsection{Proposed approach}\label{sub:proposed-approach}

\cite{schafer2011sequential} designed a tempering SMC sampler for standard
(one-level) variable selection for linear regressions, and showed it
outperformed significantly MCMC, as explained above. We adapt this approach to
our problem in three ways. First, we replace it by a waste-free SMC sampler,
following \cite{wastefreeSMC}, as waste-free SMC tends to outperform standard
SMC. Waste-free SMC amounts to resampling only a fraction of the particles,
then moving them through numerous MCMC steps, and keeping all these
intermediate. Second, we adapt the proposal mechanism within the MCMC step so
as to accommodate the constraints specific to bi-level selection (namely, that
a variable may be selected only if its group is selected).

Third, we replace the intractable marginal likelihood (obtained by integrating
out the regression coefficients) by either its LA (Laplace approximation), or
by a cheaper approximation introduced by \cite{rossell2021approximate}, called
ALA (approximate LA). The reason why ALA is particularly attractive in our
context is that it scales very well with respect to $n$ (as we explain
later). We assess in our numerical experiments the impact of  the error
introduced by ALA on the actual results. We note that \cite{schafer:thesis}
already showed in his PhD thesis that replacing the marginal likelihood by its
LA within a SMC sampler (targeting a variable selection posterior) incurs only a negligible bias. 

\subsection{Plan}\label{sub:plan}

Section~\ref{sec:model} describes the considered class
of model, the bi-level variable selection problem, and the related
notations.
Section~\ref{sec:algo} describes the proposed algorithm,
starting with a generic (waste-free) SMC sampler, and explaining how
this generic algorithm may be adapted to bi-level variable selection.
Section~\ref{sec:num} assesses (statistically and
numerically) the proposed approach through two numerical experiments,
one on simulated data and one on the public health dataset mentioned in
the introduction.

\section{Model}\label{sec:model}

\subsection{Regression model}

For the sake of concreteness, we consider the following binary regression
model, although our approach could easily be generalised to  other generalised
linear models. We suppose that we have collected a dataset $\mathcal{D} =
\left\{X,U,Z,y\right\}$ with sample size $n$, where $y\in \left\{0,1\right\}^n$
is a vector of binary responses, $X = (x_{ij}) \in \mathbb{R}^{n\times p}$,  $U
= (u_{ij}) \in \mathbb{R}^{n\times q}$, and  $Z = (z_{ij}) \in
\mathbb{R}^{n\times r}$, are design matrices that contain,  respectively, 
`individual variables',  `group variables' (both subject to variable selection
later on), and extra variables  that the user wants to include systematically
(e.g. the intercept,  socio-demographic effects such as sex, age, etc.).

Regarding the group structure, we assume that each of the $p$ variables in $X$
belongs to one (and only one) of the $q$ groups; let $g(j)$ be the group of
variable $j$. A group variable (in $U$) may represent different types of `group
effects'. For instance, in a medical application, the variables in a group $k$
may be the indicator that the patient took  a certain drug in the last six
months, and the group variable may be the  indicator that a patient took any
drug in that group in the same period.  Alternatively, these variables could be
the number of drug intakes for each drug; in that case, the group variable
would be the number of intakes of drugs in that group.  In either scenarios,
the point is to determine whether one may measure a  significant effect for
each individual variable, \emph{on top of} the group effect, or a significant
effect for its group only, or neither.  


To sum up, without variable selection, the distribution of each data point 
would be such that, for $i=1,\ldots, n$:
\begin{equation} \label{eq:lik_nosel}
 P(Y_{i}=1|\beta)=F\left(
 \sum_{j=1}^{p}\beta_{j}^{x}x_{ij}
+ \sum_{k=1}^{q}\beta_{k}^{u}u_{ik} 
+ \sum_{l=1}^{r}\beta_{l}^{z}z_{il} 
\right) 
\end{equation}
and $P(Y_i=0|\beta) = 1 - P(Y_i=1|\beta)$, 
where $\beta = (\beta^{x}, \beta^{u}, \beta^{z})$ is the vector of regression
parameters, $F$ is the link function (e.g. $F=\Phi$, the unit Gaussian CDF
for a probit model). We assign independent Gaussian priors to the regression
coefficients: $p(\beta^{z}) \sim \mathcal{N}(0_{r},
\sigma^{2}\text{I}_{r}), p(\beta^{u}) \sim \mathcal{N}(0_{q},
\sigma^{2}\text{I}_{q})$ and $p(\beta^{x}) \sim \mathcal{N}(0_{p},
\sigma^{2}\text{I}_{p})$. 

\subsection{Bi-level variable selection}

We extend our model to perform selection of groups and variables
simultaneously. Most of existing models lack flexibility as they impose only
“all-in” or “all-out” selection for variables in the same group. That is, if a group is not
selected by the model, variables belonging to this group will also not be
selected. In this work, we propose a more general approach in order to capture
sparsity at both the group and variable levels. To this end, we introduce
$\theta=(\gamma, \eta)$, a set of  two types of binary variables:  $\gamma_{k}$
indicates whether group $k$ is active ($\gamma_{k} = 1$) or not ($\gamma_{k} =
0$), and $\eta_{j}$ indicates whether individual variable $j$, which is in
group $g(j)$, is active ($\eta_{j}=1$) or not ($\eta_{j} = 0)$. We consider a
hierarchical structure such that the variable
$j$ is not selected if $\gamma_{g(j)} = 0$, that is $P(\eta_{j}=1|\gamma_{k}=0) =
0$ for $k=g(j)$. As compared to existing models, we propose to keep the flexibility of
selecting variables within a group. For example, when a group of drugs is
related to SCD, it does not necessarily mean that all drugs of this group are
related to SCD. Therefore, we may want to not only remove unimportant groups
effectively, but also identify important variables within important
groups as well. Thus, we replace  
\eqref{eq:lik_nosel} by
\begin{equation}\label{eqn:e1}
    P(Y_{i}=1|\beta,\theta)=F\left(
    \sum_{j=1}^{p}\eta_{j}\beta_{j}^{x}x_{ij}
    + \sum_{k=1}^{q}\gamma_{k}\beta_{k}^{u}u_{ik} 
    + \sum_{l=1}^{r}\beta_{l}^{z}z_{il} 
\right).
\end{equation}

Let $p(\gamma)$ be the prior density of $\gamma$, which is a product of Bernoulli
distributions with probabilities $p_{j}^\gamma$. For the predictors,
we introduce a spike-and-slab prior defined by 
\begin{equation} 
\label{eqn:e2}
P(\eta_{j}=1|\gamma) = 
\begin{cases}
    p_j^\eta & \text{if } \gamma_{g(j)} = 1 \\
    0         & \text{otherwise.}
\end{cases}
\end{equation} 
This bi-level structure implies that variable $j$ may be selected only if the
group it belongs to, $g(j)$, is selected. 

To perform Bayesian bi-level variable selection, we aim to approximating  the
(marginal) posterior distribution of $\theta =(\gamma,\eta)$, i.e.
$\pi(\theta) = p(\theta|\mathcal{D}) \propto p(\theta) L(\theta)$, where 
$p(\theta)$ is the prior described above,  
and $L(\theta)$ is the integrated likelihood obtained by integrating out
$\beta$: 
\[ L(\theta) = \int L(\beta, \theta)    p(\beta)d\beta,
\qquad L(\beta, \theta) = \left\{\prod_{i=1}^N P(Y_i=y_i|\beta,
\theta)\right\}.
\]

\section{The proposed algorithm\label{sec:algo}}

\subsection{Tempering waste-free SMC}

We propose a tempering waste-free Sequential Monte Carlo (SMC) sampler to
approximate the joint posterior distribution $\pi(\theta) =
p(\theta|\mathcal{D})$. SMC methods are iterative stochastic algorithms that
approximate a sequence of probability distributions through successive
importance sampling, resampling and Markov steps. In Bayesian modeling, this
sequence can be used to interpolate between a distribution $p(\theta)$ which is
easy to sample from (e.g. the prior distribution) and a distribution of interest
$\pi(\theta)$ which may be difficult to simulate directly (i.e. the posterior
distribution). The tempering approach in particular is based 
on a sequence of tempered distributions of the form
\begin{align*}
    \forall t \geq 1, \; \pi_{t}(\theta)
    = \frac{p(\theta)L(\theta)^{\lambda_{t}}}{Z_{t}}
\end{align*}
where $p(\theta)$ is the prior density, $L(\theta)$ the likelihood,  $Z_{t}>0$
is the normalising constant and
$0=\lambda_{0}<\lambda_{1}<\hdots<\lambda_{T}=1$ is a sequence increasing from
0 to 1. This geometric bridge smoothly interpolates between the initial
distribution $p(\theta)$ and the target distribution $\pi(\theta)\propto
p(\theta) L(\theta)$.

A typical application of such an approach is the simulation of a multimodal
distribution $\pi$. Since simulating directly from such a distribution is
difficult, we may use tempering SMC instead, to sample initially from a
distribution $p$ which covers the support of $\pi$, and to move progressively
towards $\pi$ through intermediate distributions that are progressively more
and more multimodal. In this work, we combined the tempering approach with the
waste-free SMC sampler proposed by \cite{wastefreeSMC}. The main idea of this
scheme is to resample only $M$ ancestors from the $N$ particles in the standard
SMC sampler (with $M \ll N$). Each of the ancestors is then moved $P-1$ times
through a Markov kernel $K_{t}$. The $M$ chains of length $P$ are finally put
together to form a new particle sample of size $N=MP$. Algorithm~\ref{algo:a1}
describes the corresponding algorithm for a tempering sequence. At the  final
iteration $T$ of the algorithm, one may approximate any expectation
$\mathbb{E}_{\pi}\varphi(\theta)$ with 
$\sum_{n=1}^{N}W_{T}^{n}\varphi(\theta_{T}^{n})$, where the $W_T^n$ are the
normalised weights at the final iteration $T$. 

\begin{algorithm}
    \SetKwInOut{Input}{Input}
    \SetKwInOut{Output}{Output}
    
  \Input{Prior distribution $p(\theta)$, likelihood function $\theta\rightarrow
      L(\theta)$, integers $N$, $M$, $P$ such that
  $N=MP$, sequence $0=\lambda_0<\ldots<\lambda_T=1$,
  Markov kernels $K_{t}$ that leave invariant $\pi_{t-1} \; \forall t \geq 1$}
   \For{t $\leftarrow$ 0 to T}{
   \If{$t=0$}{
      \For{n $\leftarrow$ 1 to N}{
    $\theta_{0}^{n} \sim$ $p(\theta)$}
   }
   \Else{$A_{t}^{1:M} \sim$ resample ($M,W_{t-1}^{1:N})$ (Draw IID variables
     such that $P(A_t^m=n)=W_{t-1}^n$ for $n=1,\ldots,N$)}
    \For{m $\leftarrow$ 1 to M}{
    $\tilde{\theta}_{t}^{m,1} \leftarrow \theta_{t-1}^{A_{t}^{m}}$\\
    \For{p $\leftarrow$ 2 to P}{
    $\tilde{\theta}_{t}^{m,p}\sim K_{t}(\tilde{\theta}_{t}^{m,p-1}, d\theta_{t})$ 
    }
    }
    Gather variables $\tilde{\theta}_{t}^{m,P}$ so as to form a new sample $\theta_{t}^{1:N}$\\
    \For{n $\leftarrow$ 1 to N}{
   $w_{t}^{n} \leftarrow L(\theta_{t}^{n})^{\lambda_t - \lambda_{t-1}}$ 
   }
   \For{n $\leftarrow$ 1 to N}{
   $W_{t}^{n} \leftarrow w_{t}^{n} / \sum_{m=1}^{N}w_{t}^{m}$}
}
    \caption{Tempering Waste-free SMC\label{algo:a1}}
\end{algorithm}

In practice, it is recommended to set the successive $\lambda_t$ automatically,
by choosing the next $\lambda_t$ so that the ESS (effective sample size) of the
weights equal a certain threshold. 
Another advantage of a SMC sampler such as Algorithm~\ref{algo:a1} is that it
is easy to parallelise; in particular the evaluation of the likelihood of the
$N$ particles (which is typically the bulk of the computation) may be performed
in parallel.  We refer to \cite{wastefreeSMC} for a more thorough discussion of
the advantages of SMC samplers over MCMC, and the extra advantage brought by
waste-free SMC (relative to standard SMC),  in particular the greater
robustness relative to the choice of tuning parameters such as $P$ and $M$. 

For now, there are two points that need to be addressed in order to apply
Algorithm~\ref{algo:a1} to our variable selection problem: first, we need to
design Markov kernels $K_t$ that leave invariant $\pi_{t-1}$ at time $t$, and
in particular that sample within the constrained support of $\pi_{t-1}$ in our
bi-level selection scenario (i.e. the fact that $\eta_{j}=0$ as soon as 
$\gamma_{g(j)}=0$). Second, we must find a way to evaluate, or approximate,
the marginal likelihood $L(\theta)$. These two points are discussed in the next
two sections.

\subsection{$\pi_{t-1}-$invariant kernels}

Consider a target distribution over binary vectors; that is $\pi(\gamma)$ with 
$\gamma\in \{0, 1\}^q$. Designing an efficient MCMC kernel that leaves
invariant this target is challenging. One option is to use a Gibbs kernel, or a
Metropolis kernel based on a local proposal, where only one component may be  flipped at a time. But such kernels tend to mix poorly, and to
get stuck in local modes. 

The SMC sampler of \cite{schafer2011sequential} used instead an independent
Metropolis kernel based on a global proposal of the form: 
\begin{equation}
    \label{eq:proposal_schafer}
q(\gamma) = q_1(\gamma_1) \prod_{k=1}^q q_k(\gamma_k | \gamma_{1:k-1}),
    \qquad q_k(\gamma_k=1|\gamma_{1:k-1}) = 
    \mathrm{logistic}\left(b_{kk} + \sum_{i=1}^{k-1} b_{ki} \gamma_i\right). 
\end{equation}
that is, a sequence of nested logistic regressions. 
Given the chain rule decomposition above, it is easy to sample
from this proposal distribution. In order to ensure that the resulting
independent Metropolis sampler mixes well (and in particular that the
acceptance rate is high), one needs to ensure that the proposal is as close as
possible to the target. To ensure this, \cite{schafer2011sequential} set the
parameters $b_{ji}$ to the maximum likelihood estimators of the corresponding
logistic regressions, based on the current (weighted) particle sample. The numerical
experiments of \cite{schafer2011sequential} show that a SMC sampler based on
such  global (properly calibrated) Metropolis steps may outperform
significantly local MCMC chains.

Since \cite{schafer2011sequential} considered standard (one-level) variable
selection, they did not have to deal with constrained distribution (i.e. each
vector $\gamma\in\{0, 1\}^p$ has positive probability). We adapt their approach
to bi-level variable selection as follows. First, we extend the proposal
in~\eqref{eq:proposal_schafer} as follows:
\begin{equation}
    \label{eq:proposal_bi}
 q(\theta) 
    = q(\gamma, \eta) 
    = q_1(\gamma_1) \prod_{k=1}^q q_k(\gamma_k | \gamma_{1:k-1})
    \prod_{j=1}^p q_j(\eta_j|\gamma_{g(j)}). 
\end{equation}
where the conditional distributions of the $\gamma_j'$s are set in the same way
as in~\eqref{eq:proposal_schafer}.  Second, we set the conditional proposals of
the $\eta_j$ as follows: 
\[
    q_j(\eta_j=1|\gamma_{g(j)}) = 
    \begin{cases}
        c_j & \text{if } \gamma_{g(j)} = 1 \\
        0   & \text{otherwise}
    \end{cases}
\]
where $c_j\in[0, 1]$ is a tuning parameter. We calibrate the $c_j$'s in the
same way as for the coefficients $b_{ji}$ in~\eqref{eq:proposal_schafer}: by
maximum likelihood estimation on the current particle sample. 

This proposal respects the constraint that $\eta_j$ must be zero as soon as
$\gamma_{g(j)}=0$. It is basic, and may be extended by correlating the
$\eta_j'$s in the same group through a nested logistic regression of the same
form as for the $\gamma_k$. In practice however, we did not observe much
benefit in doing so, and stuck to this basic structure.
Algorithm~\ref{alg:metropolis} summaries how one may implement the considered
type of Metropolis kernels. 

\begin{algorithm}
    \SetKwInOut{Input}{Input}
    \SetKwInOut{Output}{Output}
    \Input{$\theta=(\gamma, \eta)$, tuning parameters $(b_{ji})$ and $(c_j)$
    (estimated from the current particle sample).}
   \Output{A sample from $K_t(\theta, d\theta')$, where $K_t$ leaves invariant
   $\pi_{t-1}$.}
   $\theta^p \sim q(\theta)$ (as defined in~\eqref{eq:proposal_bi})\\
   $u\sim \mathrm{Uniform}[0, 1]$ \\
   \If{ $u \leq \pi_{t-1}(\theta^p) q(\theta) / \pi_{t-1}(\theta) q(\theta^p)$ }
   {\Return{$\theta^p$}}
   \Else{\Return{$\theta$}}

    \caption{Independent Metropolis kernel used to move the particles within
    Algorithm 1 at time $t$\label{alg:metropolis}}
\end{algorithm}

\subsection{Approximation of the marginal likelihood}

The marginal likelihood $L(\theta) = \int L(\beta,\theta) p(\beta) d\beta$ is typically
intractable (unless one considers a linear Gaussian regression model). A
popular approximation to this quantity is the Laplace approximation (LA), which
amounts to Taylor expanding the log of the integrand around its mode. Let
$\beta_\theta$ denote the vector made of the components $\beta_i$ such  
that $\theta_i=1$, 
$h_\theta(\beta_\theta) = - \log \{L(\beta, \theta) p(\beta)\}$, and
$\hat{\beta}_\theta=\arg\min_{\beta_\theta}h_\theta(\beta_\theta)$ (i.e. 
the MAP estimator given $\theta$), then
\begin{align*} 
    \log L(\theta) 
    & = \log \int \exp\left\{ - h_\theta(\beta_\theta) \right\} d\beta_\theta\\
    & \approx - h_\theta(\hat\beta_\theta) 
    + \log \int \exp\left\{ -\frac{1}{2} (\beta_\theta-\hat\beta_\theta)^T
    \hat{H}_\theta (\beta_\theta - \hat\beta_\theta) \right\} d\beta_\theta \\
    & = - h_\theta(\hat\beta_\theta) 
    + \frac{d_\theta}{2}\log 2\pi - \frac{1}{2} \log |\hat{H}_\theta|
\end{align*}
where $|\hat{H}_\theta|$ is the determinant of the Hessian of function
$\beta_\theta \rightarrow h_\theta(\beta_\theta)$ at
$\beta_\theta=\hat\beta_\theta$,
and $d_\theta=\dim\beta_\theta$. 

\cite{schafer:thesis} in his thesis gave numerical evidence than replacing the
marginal likelihood with its Laplace approximation, within a SMC sampler for
standard (one-level) variable selection, works well, in the sense that it 
 leads to a negligible error (for approximating the posterior of $\theta$).
On the other hand, computing the Laplace approximation for many simulated
$\theta-$values is expensive; for each $\theta$, one needs to run a
Newton-Raphson optimiser to obtain $\hat\beta_\theta$ and $\hat{H}_\theta$. 
Furthermore these operations have complexity $\mathcal{O}(n)$ in the sample
size, and $\mathcal{O}(d_\theta^3)$ in the dimension. 

\cite{rossell2021approximate} proposed a cheaper approximation, based on a
Taylor expansion similar to Laplace, but around zero. Let
$\mathbf{0}_\theta$ denote a vector of zeros of the same dimension as
$\beta_\theta$, then, the ALA (approximate Laplace approximation) is
\begin{align*} 
    \log L(\theta) 
    & \approx - h_\theta(\mathbf{0}_\theta) 
    + \log \int \exp\left\{ - \beta_\theta^T g_\theta -\frac{1}{2} \beta_{\theta}^T
        H_\theta \beta_\theta  \right\} d\beta_\theta \\
    & = - h_\theta(\mathbf{0}_\theta) 
    + \frac 1 2 g_\theta^T H_\theta^{-1} g_\theta 
    + \frac{d_\theta}{2}\log 2\pi - \frac{1}{2} \log |H_\theta|
\end{align*}
where $g_\theta$ and $H_\theta$ denote respectively the gradient and Hessian of function
$\beta\rightarrow h_\theta(\beta_\theta)$ at point $\beta_\theta=\mathbf{0}_\theta$. Note that in
practice, one simply need to compute the gradient $g$ and Hessian $H$ of minus
log-likelihood at zero for the \emph{full} model (i.e. $\theta$ is a vector of ones,
all variables are included), to obtain $g_\theta$ and $H_\theta$ (e.g.
$g_\theta$ contains the components $i$ of $g$ such that $\theta(i)=1$, and
$H_\theta$ is defined similarly). 

Once quantities $g$ and $H$ have been computed in a preliminary step, the
computation of ALA is $\mathcal{O}(1)$ in the sample size $n$. 
Its complexity remains cubic in the dimension, because of the determinant,
however.
\cite{rossell2021approximate} make it clear that ALA is not
a consistent (in $n$) approximation of the marginal likelihood; they mention
that it tends to be biased against truly active variables. That is, it tends to
under-estimate the posterior probability that an active variable should be
included. We refer to \cite{rossell2021approximate} for more discussion on this
matter. 

Still, ALA remains particularly attractive in our context, as our SMC sampler
must perform many evaluations of the marginal likelihood. We will assess
the impact of the approximation error of ALA by comparing two waste-free SMC
samplers, one based on LA, and one based on ALA.

\section{Numerical experiments}
\label{sec:num}

As explained above, our goal in this section is to assess numerically the
performance of our tempering waste-free SMC sampler for bi-variable selection,
when the marginal likelihood is evaluated through either LA or ALA. We take the
number of particles to be $N=25, 000$, and set $M=125$, $P=200$. Our algorithm
was implemented using  the particles Python library (see 
\url{https://github.com/nchopin/particles}).
The results were obtained using a server with 64 Gb RAM and 8 cores.

\subsection{Simulated data}

We simulate data from our model (using the probit link function), using  $g=5$
groups, $r=5$ systematically included covariates, a varying number $p$ of
individual variables, and a varying sample size $n$; see below.  The rows of
the design matrices $X$, $U$, and $Z$  are sampled independently from a
Gaussian distribution $N(0, \Sigma)$, where $\Sigma_{ii}=1$, and
$\Sigma_{ij}=0.5$. The corresponding regression parameters are set to 
$\beta^{z}=(0,0,1,1,1)$, $\beta^{u}=(0,0,1,1,1)$ and the components of
$\beta^{x}$ are all set to zero, except for the last variable of each active
group, where it is set to one. 


In a first scenario, we set $p=50$ and let $n$ vary from 100 to $2,500$; while
in a second scenario we fix $n=1,500$ and let  $p$ vary from 10 to 250. 
We run our algorithm 10 times and uses the empirical standard deviation to draw 
confidence intervals.

\begin{figure}
\includegraphics[scale=.3]{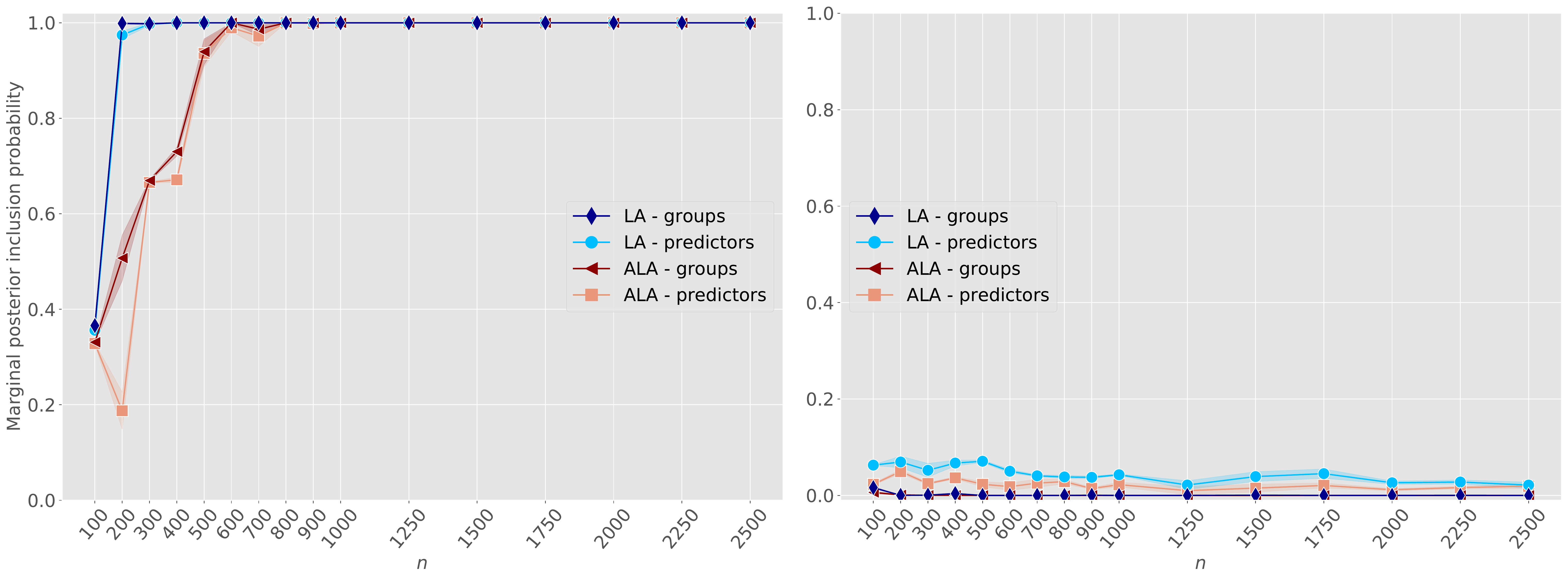}
\caption{\label{fig:simulated_data_scenario1}Comparison 
of ALA and LA for posterior inclusion probabilities of
groups and predictors when $n$ varies from 100 to $2,500$, with $p=50$. Left:
average posterior inclusion probabilities for truly active variables. Right:
average posterior inclusion for truly inactive variables.}
\end{figure}

\begin{figure}
\includegraphics[scale=.3]{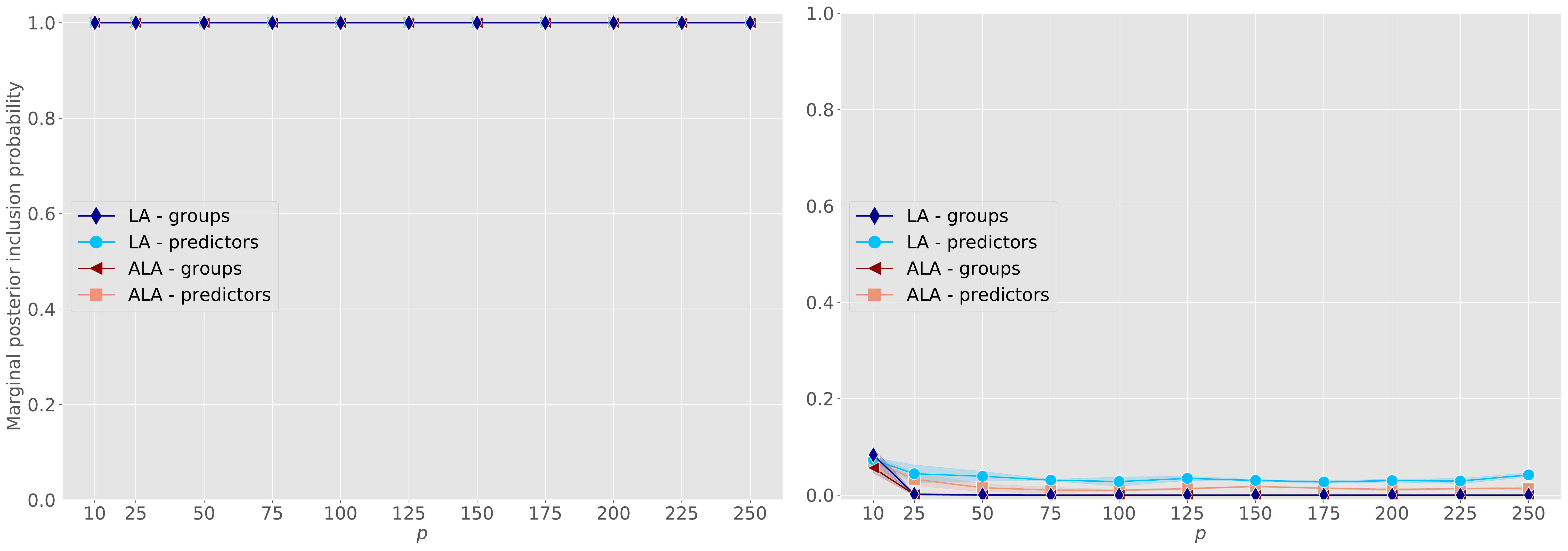}
\caption{\label{fig:simulated_data_scenario2}Comparison 
of ALA and LA for posterior inclusion probabilities of
groups and predictors when $p$ varies from 10 to 25, with $n=1,500$. Left:
average posterior inclusion probabilities for truly active variables. Right:
average posterior inclusion for truly inactive variables.}
\end{figure}

Figure~\ref{fig:simulated_data_scenario1} summarizes the results from the first scenario. 
Both LA and ALA discriminate properly truly active from inactive groups and
variables when $n$ is large enough. However, LA assigns larger inclusion
probabilities for truly variables when $n \leq 500$.
Figure~\ref{fig:simulated_data_scenario2} summarizes the
results for the $n=1,500$ case, when $p$ varies from 10 to 25. LA and ALA
performed equally and provided accurate estimates both for groups and
variables.

\begin{figure}
\includegraphics[scale=.3]{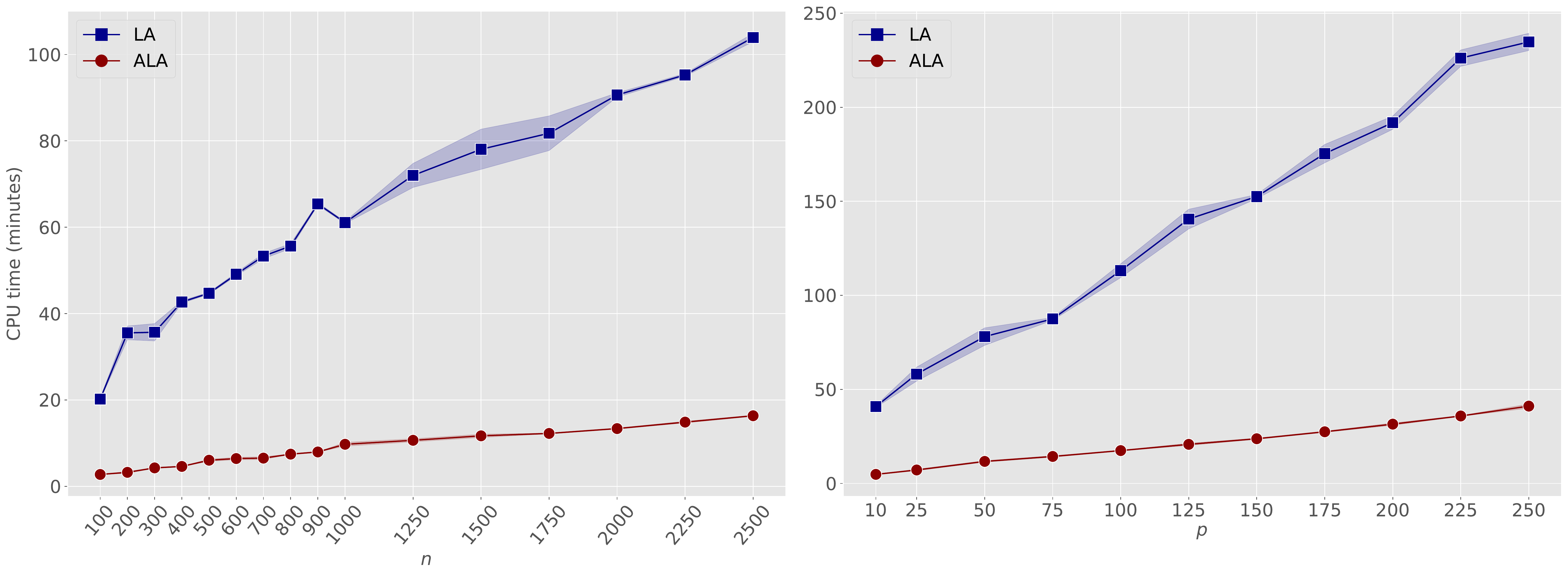}
\caption{\label{fig:sim_cpu1}Comparison of ALA and LA for run time of waste-free SMC. Left: average run time when $n$ varies from 100 to $2,500$ ($p=50$). Right: average run time when $p$ varies from 10 to $25$ ($n=1,500$). }
\end{figure}

Figure~\ref{fig:sim_cpu1} compares the performance of ALA and LA in terms of
computation time in both scenarios. ALA significantly reduces run times
compared to LA, especially for larger $n$ (mean run time = 16 min for ALA
\textit{vs.} 102 min for LA when $n=2,500$ and $p=50$) and $p$ (mean run time =
39 min for ALA \textit{vs.} 330 min for LA when $p=250$ and $n=1,500$). 
It is interesting to note that the CPU time still grows with $n$ with ALA,
although the computation of ALA is independent of $n$. The likely explanation
is that when $n$ grows, the prior and the posterior differ more markedly, and
thus more intermediate tempering distributions are required to bridge between
the two. Still, the dependence on $n$ of the CPU time remains mild compared to
the LA-based sampler. 

To sum up, one observes that ALA considerably reduces the CPU time of the
sampler, in particular for large $n$ (sample size) and $p$ (number of
variables). In return, as expected ALA tends to under-estimate the probability of
inclusion of active variables, at least for $n$ not sufficiently large.  

\subsection{Bi-level selection on the French National Healthcare Insurance database}

To examine the performance of our SMC sampler on a big dataset, we
study which factors are associated to sudden cardiac death (SCD) in a French
epidemiological study. Sudden cardiac death is an unexpected death due to
cardiac causes that occurs in a short time period (generally within 1 hour of
symptom onset) in a person with known or unknown cardiac disease. Despite
progress in epidemiology, clinical profiling and interventions, it remains a
major public health problem worldwide, accounting for 10 to 20\% of deaths in
industrialised countries. The annual incidence of SCD is estimated 180,000 to
450,000 in the United States (\cite{Kong2011}) and 275,000 in Europe (\cite{Empana2022}). The prognosis is terrible,
with less than 10\% surviving to hospital discharge, and significant functional
and cognitive disabilities often persist among those who survive (\cite{Bougouin2014}). Therefore,
identification of persons with an elevated risk of SCD is highly relevant from
a clinical and public health perspective. 

In this study, we implement bi-level variable selection to identify outpatient drugs
and hospital diagnoses that could help to enhance risk prediction performance
of SCD over many potential risk factors collected from electronic health records. We analyse the medical
trajectories of $n_{\mathrm{cases}} = 23,958$ cases of SCD collected between 2016 and 2020
throughout the Paris Sudden Death Expertise Center (\cite{Bougouin2014}), and $n_{\mathrm{controls}} = 23,958$ controls
sampled from the French general population. Cases and controls were matched
with age, sex and residence area. 

For the $n = n_{\mathrm{cases}} + n_{\mathrm{controls}} = 47,916$ individuals, we
collected data from the French National Health Insurance (SNDS) database, which
manages all reimbursements of healthcare for people affiliated to a health
insurance scheme in France. It provides information on all healthcare expenses,
on an individual level, including visits, procedures and reimbursed
drugs relative to outpatient medical care claims, information from hospital
discharge summaries and chronic conditions. Data acquisition is permanent,
from birth to death, irrespective of wealth, age, or work status, resulting in
one of the largest electronic health records databases in the world. The SNDS database has been described in detail previously and has been used to conduct multiple studies in cardiovascular
epidemiology (\cite{Piot2022}). More details are available at \url{https://www.health-data-hub.fr/}.\\

\begin{figure}[H]
\includegraphics[scale=.5]{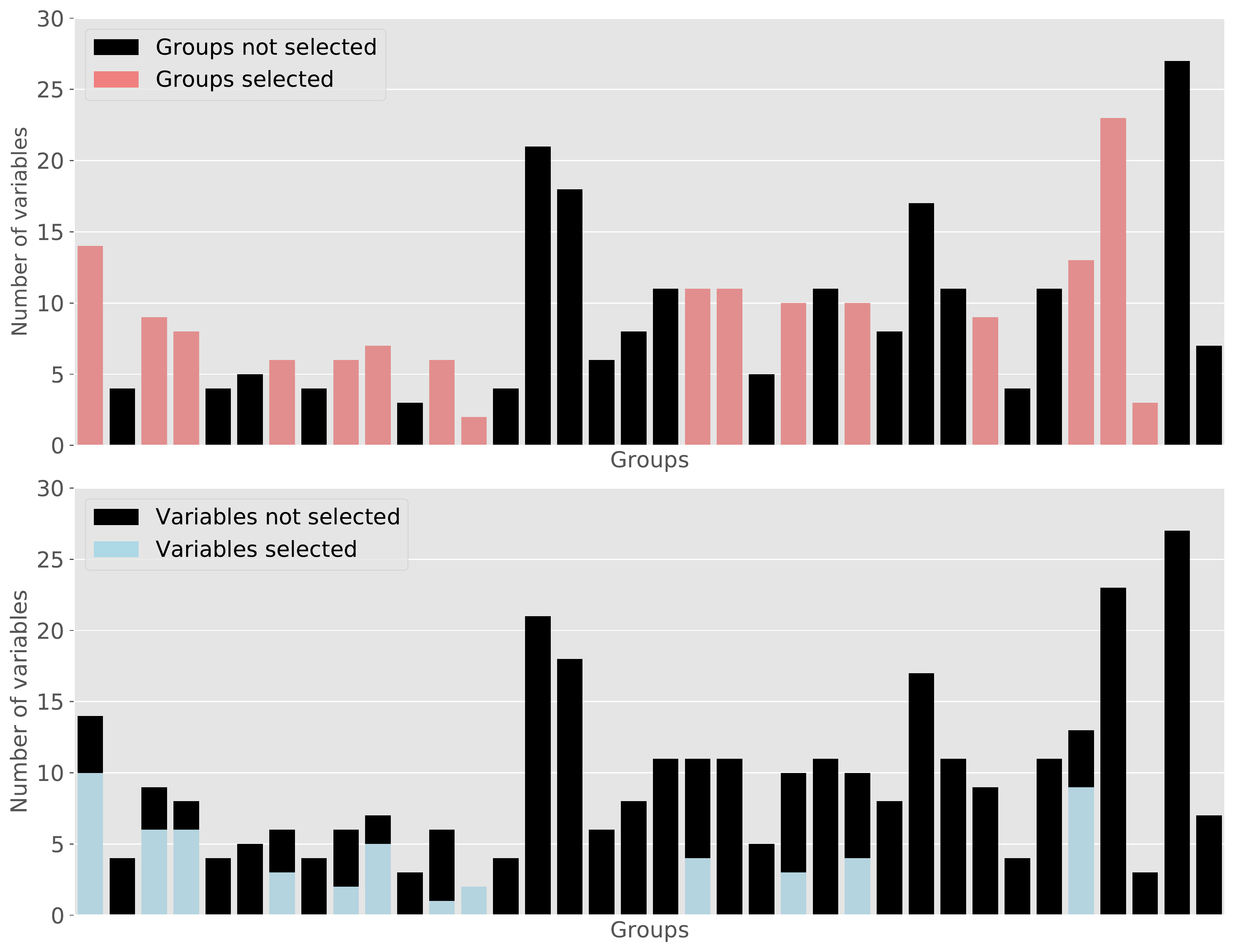}
\caption{\label{fig:selected_variables_real_data}Groups and predictors selected by the ALA-based SMC sampler. Top: selection of groups. Bottom: selection of predictors.}
\end{figure}

\begin{figure}[H]
\includegraphics[scale=.4]{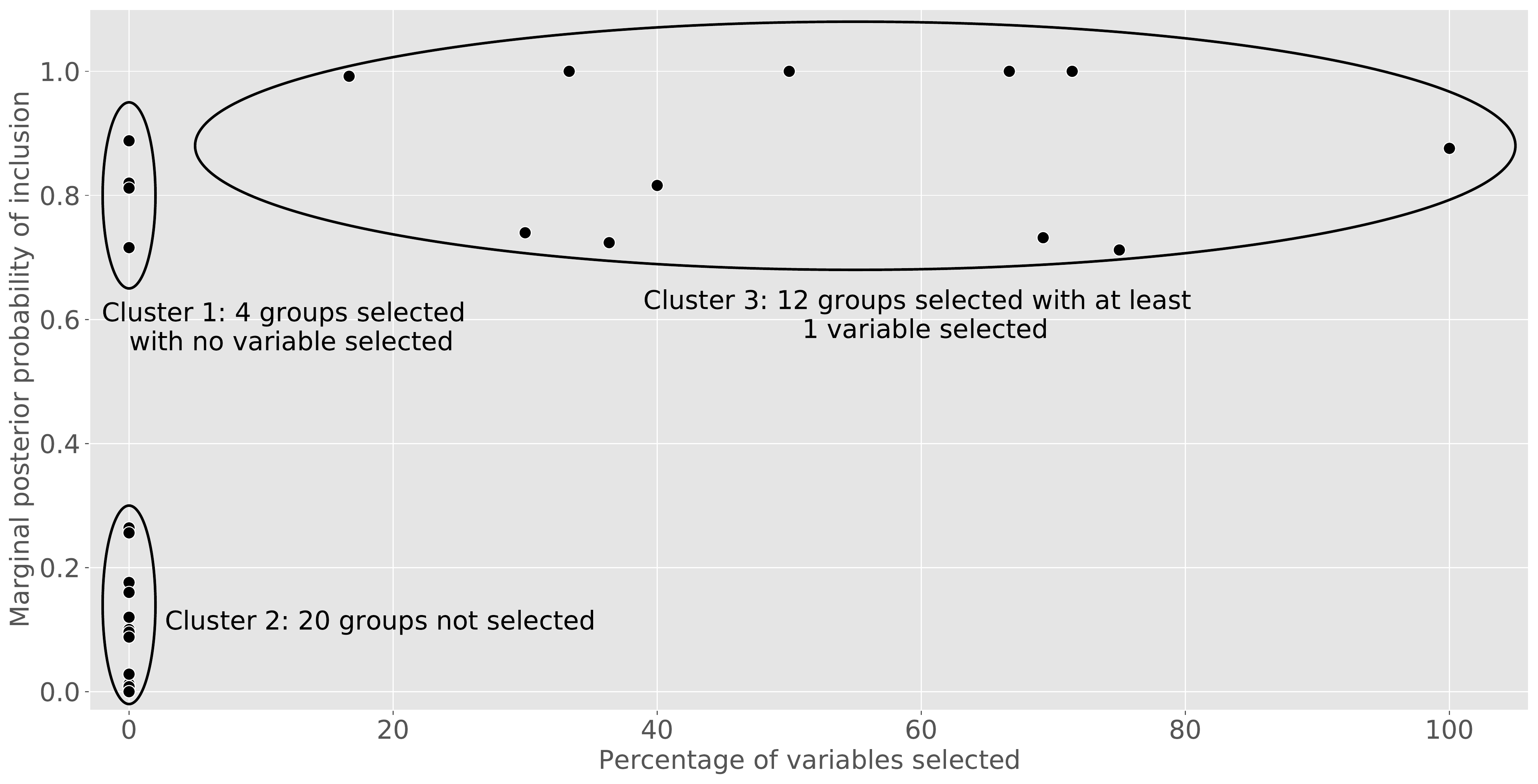}
\caption{\label{fig:scenarios_real_data}Bi-level variable selection scheme proposed by the ALA-based SMC sampler}
\end{figure}

We collected all outpatient drugs and hospital diagnoses that occurred up to 5 years before SCD; in this way we obtained $q = 36$ groups and $p = 337$ binary variables (0/1 whether the individual took a particular drug in the last 5 years, or a drug in the corresponding group). In the 36 groups, the minimum number of variables observed is 2 and the maximum is 27. No external variables were included in the study ($r=0$). Figures~\ref{fig:selected_variables_real_data} and \ref{fig:scenarios_real_data} summarise the results of our ALA-based SMC sampler in terms of variable (and group) selection. We evaluate groups and variables selected by our model by comparing them with those described in the medical literature related to SCD. Overall, 16 out of 36 groups and 55 out of 337 variables are selected (Figure~\ref{fig:selected_variables_real_data}). Our bi-level variable selection scheme allows for a more flexible structure than "all-in all-out" methods and identifies 3 different "clusters" represented in Figure~\ref{fig:scenarios_real_data}. 

In the first cluster (located in the upper left corner), 4 groups of hospital diagnoses are selected without any variable included. These groups correspond to diseases of the eye ($\pi(\gamma_{k}=1)=0.82$), diagnoses related to pregnancy, childbirth and the puerperium ($\pi(\gamma_{k}=1)=0.82$), injury and poisoning ($\pi(\gamma_{k}=1)=0.72$) and diagnoses for other special purposes ($\pi(\gamma_{k}=1)=0.89$). They are selected with high marginal posterior probabilities of inclusion, although none of their 46 corresponding variables are selected. This result suggests therefore that only global relationships exist between these groups and SCD, with no any precise effect of diseases or treatments. 

In the second cluster (located in the lower left corner), 20 groups are not selected, as well as their 189 corresponding variables. They include diverse subgroups of diseases and treatments. 

In the third cluster (located in the upper right corner), 12 groups are selected with at least 1 variable included. Among them, 3 well known groups of risk factors of SCD are identified. First, diseases and drugs associated to the cardiovascular system are selected (with $\pi(\gamma_{k}=1)=0.74$ and $\pi(\gamma_{k}=1)=1$ respectively), including 9 out of 19 variables. This result was expected, as cardiovascular conditions are known to be the most common pathology under SCD. Second, diseases and drugs related to the nervous system are selected (with $\pi(\gamma_{k}=1)=0.72$ and $\pi(\gamma_{k}=1)=1$ respectively), including 9 out of 18 variables. Several studies have suggested relationships between diseases of the nervous system and SCD (\cite{Japundzic2018}). Indeed, some neurological disorders can cause damage to the heart and blood vessels (such as stroke or brain injury) or arrhythmia (such as epilepsy), increasing the risk of SCD. There are also neurological conditions that can cause SCD directly, such as long QT or Brugada syndromes, which affect the electrical activity of the heart. Third, a group related to treatments of the respiratory system is selected. A number of studies have also addressed the relationship between respiratory disorders and SCD. In particular, 
cumulating evidence associates chronic obstructive pulmonary diseases with an increased risk of SCD both in cardiovascular patient groups and in community-based studies, independent of cardiovascular risk profile (\cite{VandenBerg2016}). 

\begin{figure}
\includegraphics[scale=.4]{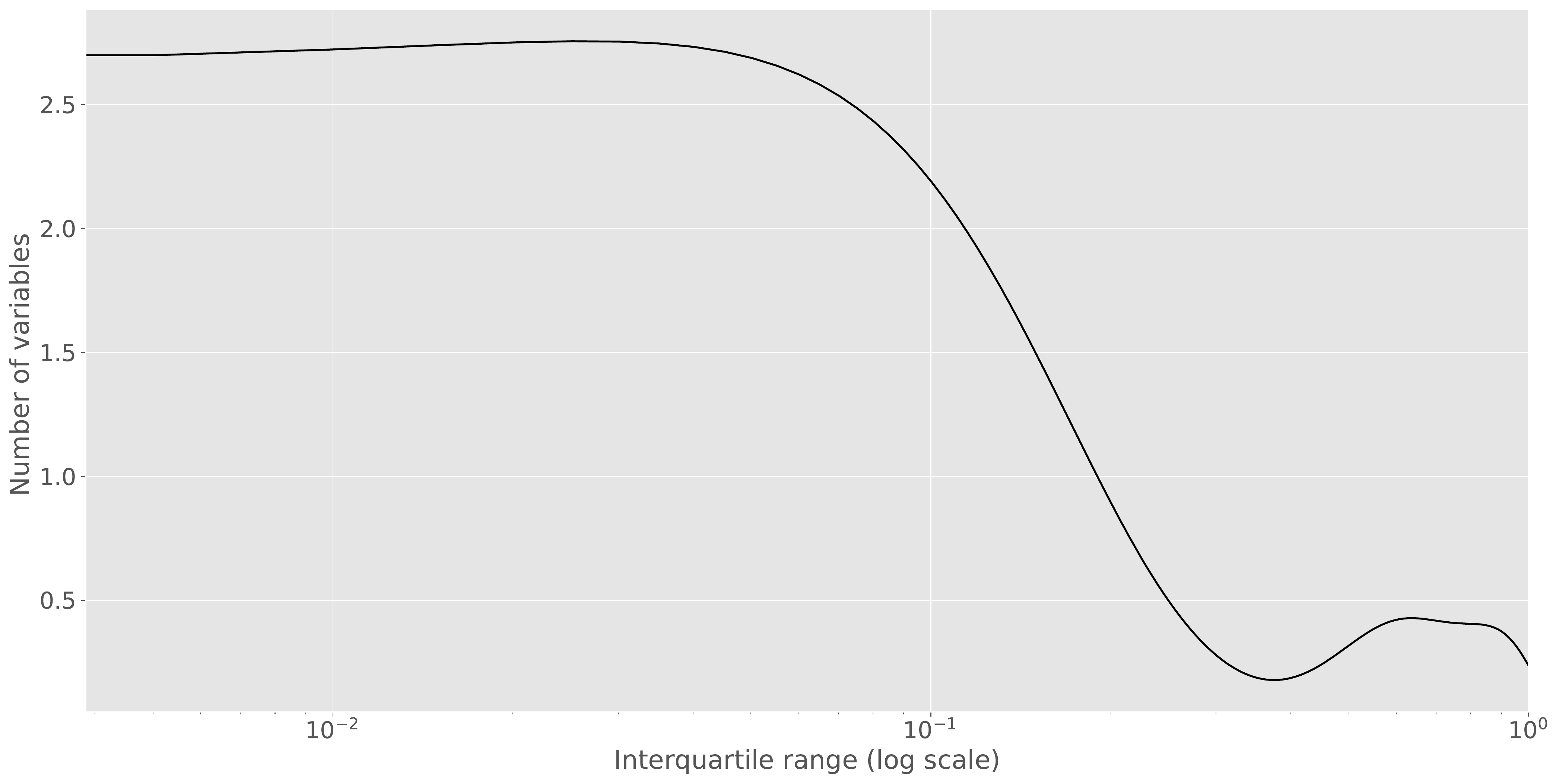}
\caption{\label{fig:stability_real_data}Kernel density estimate of the interquartile range (log scale) of the marginal posterior inclusion probabilities (variables) for the ALA-based SMC sampler.}
\end{figure}

We ran our ALA-based SMC samplers 10 times to assess its numerical stability. 
Figure~\ref{fig:stability_real_data} describes the interquartile range of the marginal posterior probabilities of inclusion for variables.  The mean run time was 61.8 hours (totalling to 7 days of total CPU time). We also launched 10 executions of our LA-based SMC sampler, but these executions had not completed after 30 days. We can see that, for this particular dataset, using ALA becomes crucial to make the approach usable for practitioners. 

\section{Conclusion}

Our bi-level variable selection approach based on a waste-free SMC sampler and the ALA approximation offers reliable performance for large-scale datasets within a reasonable computation time. Furthermore, our approach is more flexible  than most of existing schemes, which impose only “all-in” or “all-out” selection for variables in the same group. This work could be therefore helpful in a wide range of applications, such as biomedical studies, where standard approaches provide information which may be difficult for physicians to interpret. 

\bibliographystyle{apalike}
\bibliography{complete} 

\end{document}